\documentclass[12pt]{iopart}

\usepackage{iopams} 
\usepackage{natbib}
\usepackage{hyperref}
\usepackage{graphicx}
\usepackage{xcolor}
\begin{document}

\title[Bayesian Reconstruction of Eccentric BBH Signals]{Bayesian Reconstruction of Gravitational-wave Signals from Binary Black Holes with Nonzero Eccentricities}

\author{Gergely D\'alya$^1$, Peter Raffai$^{1,2}$, Bence B\'ecsy$^{3}$}
\address{$^1$ Institute of Physics, E\"otv\"os University, 1117 Budapest, Hungary}
\address{$^2$ MTA-ELTE Extragalactic Astrophysics Research Group, 1117 Budapest, Hungary}
\address{$^3$ Montana State University, Bozeman, MT 59717, USA}
\ead{dalyag@caesar.elte.hu}

\vspace{10pt}
\begin{indented}
\item[]Received October 2020
\end{indented}

\begin{abstract}
We present a comprehensive study on how well gravitational-wave signals of binary black holes with nonzero eccentricities can be recovered with state of the art model-independent waveform reconstruction and parameter estimation techniques. For this we use BayesWave, a Bayesian algorithm used by the LIGO-Virgo Collaboration for unmodeled reconstructions of signal waveforms and parameters. We used two different waveform models to produce simulated signals of binary black holes with eccentric orbits and embed them in samples of simulated noise of design-sensitivity Advanced LIGO detectors. We studied the network overlaps and point estimates of central moments of signal waveforms recovered by BayesWave as a function of $e$, the eccentricity of the binary at 8 Hz orbital frequency. BayesWave recovers signals of near-circular ($e\lesssim0.2$) and highly eccentric ($e\gtrsim0.7$) binaries with network overlaps similar to that of circular ($e=0$) ones, however it produces lower network overlaps for binaries with $e\in[0.2,0.7]$. Estimation errors on central frequencies and bandwidths (measured relative to bandwidths) are nearly independent from $e$, while estimation errors on central times and durations (measured relative to durations) increase and decrease with $e$ above $e\gtrsim0.5$, respectively. We also tested how BayesWave performs when reconstructions are carried out using generalized wavelets with linear frequency evolution (chirplets) instead of sine-Gaussian wavelets. We have found that network overlaps improve by $\sim 10-20$ percent when chirplets are used, and the improvement is the highest at low ($e<0.5$) eccentricities. There is however no significant change in the estimation errors of central moments when the chirplet base is used.
\end{abstract}

%
\vspace{2pc}
\vspace*{2pc}
\submitto{\CQG}
%
%
%

\section{Introduction} \label{sec:intro}

The network of the Advanced LIGO (aLIGO, see \citealt{2015CQGra..32g4001L}) and Advanced Virgo (AdV, see \citealt{2015CQGra..32b4001A}) gravitational-wave (GW) detectors achieved the first direct detection of GWs in 2015 \citep{2016PhRvX...6d1015A}. During its first and second observing runs, the network observed a total of 11 GW signals, all but one originating from coalescing binary black holes (BBHs) \citep{2018arXiv181112907T}. These signals were found by template-based search methods that assume zero orbital eccentricities (see e.g.\ \citealt{2016CQGra..33u5004U}), and based on the method outlined in \cite{2018PhRvD..98h3028L}, \cite{2019MNRAS.490.5210R} also showed that all of these detections are consistent with having originated from binaries in circular orbits.

Observations of BBHs with nonzero orbital eccentricities should help breaking degeneracies in measuring source parameters with GWs (see e.g.\ \citealt{2018PhRvD..97b4031H}) and could shed light on key questions about possible formation channels of BBHs (see e.g.\ \citealt{2019MNRAS.486..570T}). Such systems can form through binary stellar evolution (see e.g.\ \citealt{2016Natur.534..512B}) or through dynamical processes (see e.g.\ \citealt{2014ApJ...781...45A}, \citealt{2016PhRvD..93h4029R}, \citealt{2018ApJ...855..124S}). GW emission shrinks and circularizes the orbits of BBHs on long timescales, which can lead to negligible eccentricities when the central frequency of the GW emission reaches the nominal low-frequency limit of aLIGO, i.e.\ 10 Hz \citep{2016ApJ...830L..18B} (we will denote this eccentricity as $e_{10}$ to distinguish it from $e$, the eccentricity of the binary at the time the orbital frequency is 8 Hz, which we used as the initial eccentricity of all our simulated BBHs). However some processes such as gravitational capture through GW emission in dense stellar systems \citep{2018ApJ...860....5G} or the Kozai-Lidov mechanism in hierarchical triple systems \citep{2012ApJ...757...27A} can produce BBHs with $e_{10} \gtrsim 0.1$. As \cite{2019MNRAS.486..570T} pointed out, constraining the proportions of eccentric BBHs formed through the two main formation channels could be feasible by aLIGO detections of their GWs.



Template-based GW searches by the LIGO-Virgo Collaboration currently do not include templates of binaries with nonzero eccentricities \citep{2018arXiv181112907T}. However it is shown that using only circular templates results in a significant loss in detection rate for binaries having $e_{10} \gtrsim 0.1$ (see e.g.\ \citealt{2010PhRvD..81b4007B}, \citealt{2018CQGra..35w5006M}). Searches for generic GW transients (called \textit{bursts}, see e.g.\ \citealt{2018LRR....21....3A}) that look for coincident excess power in the strain data of multiple GW detectors (e.g.\ \citealt{2012PhRvD..85l2007A}) have the potential to find BBH signals with $e_{10}>0$. One such algorithm is the Coherent WaveBurst (cWB, see \citealt{2005PhRvD..72l2002K}, \citealt{2008CQGra..25k4029K} and \citealt{2020arXiv200612604D}), which is already being used by the LIGO-Virgo Collaboration in searches for eccentric BBHs (see e.g.\ \citealt{2019ApJ...883..149A}).

BayesWave (BW, see \citealt{2015CQGra..32m5012C}, \citealt{2015PhRvD..91h4034L} and \citealt{2020arXiv201109494C}) has been used as a follow-up waveform reconstruction and parameter estimation (PE) tool on detection candidates provided by the cWB in the first three observing runs. BW works within the framework of Bayesian statistics and uses either sine-Gaussian wavelets or "chirplets" (i.e.\ modified sine-Gaussian wavelets with linear frequency evolution, see \citealt{Chirplet}) as basis functions to reconstruct a signal \citep{2018PhRvD..97j4057M}. As it is shown in e.g.\ \citealt{2016PhRvD..94d4050L} and \citealt{2016PhRvD..93b2002K}, BW is able to effectively distinguish between real astrophysical signals and non-Gaussian noise artifacts (called \textit{glitches}, see \citealt{2019arXiv190811170T}).

\cite{2017ApJ...839...15B} provides a comprehensive multi-aspect study on the performance of BW in estimating parameters of GW bursts with four different morphologies (sine-Gaussians, Gaussians, white-noise bursts and circular BBH signals). In this paper we extend this study to BBH signals with $e>0$, focusing on two aspects of BW's performance: (i) waveform reconstruction and (ii) estimation of model-independent waveform parameters. We also quantify the difference of $e>0$ BBH signal reconstructions between using a sine-Gaussian wavelet base and a chirplet base.

This paper is organized as follows. In Section 2, we describe the methods we used for generating simulated signals and noise samples, and the methods we used to characterize the performance of BW in waveform reconstructions and PE. In Section 3, we present results of our
analyses regarding waveform reconstruction and PE, as well as the comparison of BW's performance for a sine-Gaussian and a chirplet base. We summarize our findings and draw conclusions in Section 4.

\section{Methods} \label{sec:methods}

We tested the performance of BW with mock aLIGO noise samples and simulated BBH signals added to them. We considered a two-detector network consisting of the two aLIGO detectors at Hanford, WA (denoted by H1), and at Livingston, LA (denoted by L1). As \cite{2019arXiv190811170T} notes, loud glitches and GW bursts contribute to non-stationary and non-Gaussian features, but away from these transient disturbances the LIGO-Virgo data can be approximated as being stationary and Gaussian. As glitches typically occur on the order of once per minute \citep{2019arXiv190811170T}, a coincidence in both time and frequency between a glitch and a BBH signal with a duration of less than a second is relatively unlikely. Data containing signals and glitches in the same time-frequency volume can be pre-processed through a glitch-subtraction procedure based on BW (see \citealt{2018PhRvD..98h4016P}) prior to being analyzed, without compromising the signal itself. This technique has already been used for the second Gravitational-Wave Transient Catalog, see \cite{2020arXiv201014527A}. Therefore as our mock noise, we simulated stationary, Gaussian noise samples with an amplitude spectral density resembling the design sensitivity curve of aLIGO \citep{Barsottietal}. We used the 1.0.5 version\footnote{The BW version we used is publicly accessible at: \\\url{https://git.ligo.org/lscsoft/bayeswave/-/tags/v1.0.5}} of BW in all our reconstructions. 

We used two different algorithms to create simulated BBH waveforms with nonzero eccentricities. The \texttt{EccentricTD} waveform generator \citep{2016PhRvD..93f4031T} was created by incorporating orbital eccentricity into the quasi-circular time-domain 2PN-accurate \texttt{TaylorT4} approximant\footnote{The  post-Newtonian  approximation  is  one  in  which  the  field equations of general relativity  are  solved  assuming  small  velocities  and  weak  gravitational fields in an expansion in powers of ($v/c$), where $v$ is the orbital velocity and $c$ is the speed of light.  By $n$PN order we mean an expansion to order $(v/c)^{2n}$. For more details see \cite{2006LRR.....9....4B}.} (see e.g.\ \citealt{2009PhRvD..80h4043B}). The waveform models only the inspiral phase of the binary evolution and does not include BH spins. The authors of \cite{2016PhRvD..93f4031T} point out that the \texttt{EccentricTD} approximant should be accurate and efficient to handle $e<0.9$ initial orbital eccentricities. The other waveform generator (which we will refer to by the name \texttt{IMR}), described in \cite{2013PhRvD..87d3004E}, simulates full inspiral-merger-ringdown waveforms of BBHs with nonzero eccentricities, and include generic spin configurations. As the authors of \cite{2013PhRvD..87d3004E} claim, the \texttt{IMR} waveform generator is adequate to supply mock signals to explore the performance of existing LIGO searches, however it is not sufficiently accurate to generate a matched-filter template bank (similar to the one presented in \citealt{2016PhRvD..93l2003A}). The usage of this waveform generator is further justified by the fact that in \cite{2019ApJ...883..149A} it was used by the LIGO-Virgo Collaboration to characterize the sensitivity of the search for eccentric BBH mergers with the cWB algorithm during the first and second observing runs.

\begin{figure*}
\begin{center}
\includegraphics[scale=0.3,angle=0]{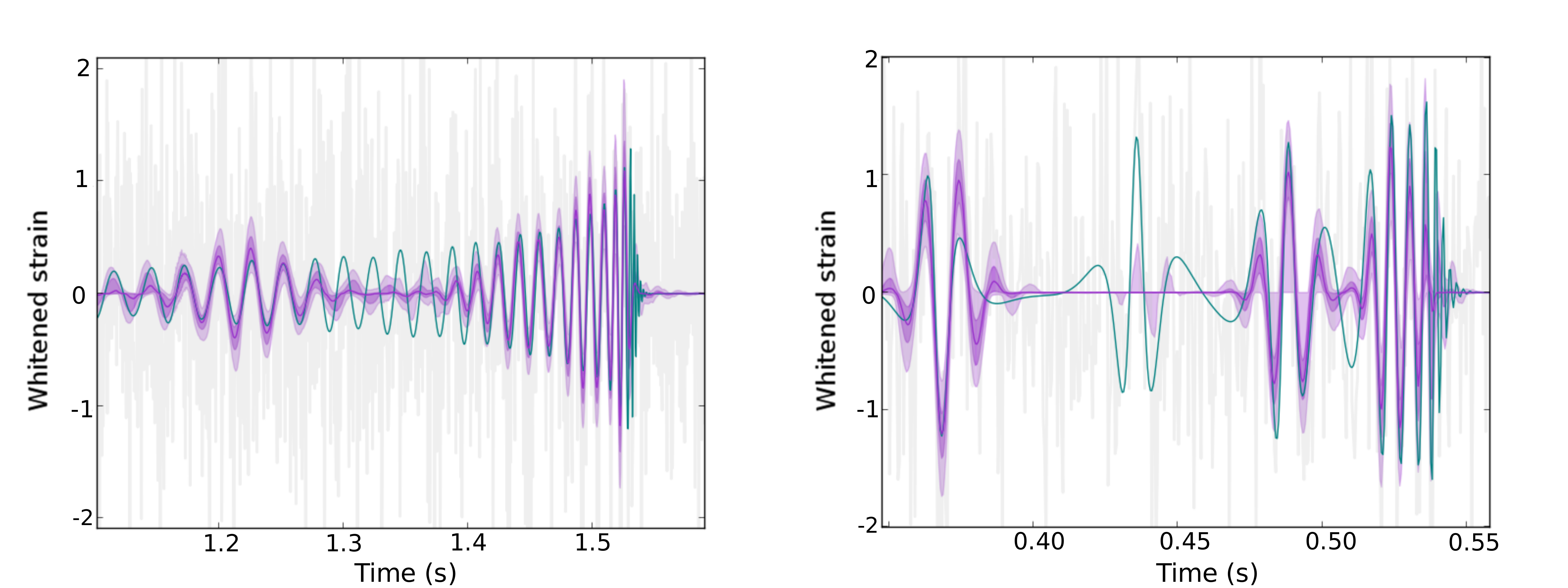}
\includegraphics[scale=0.3,angle=0]{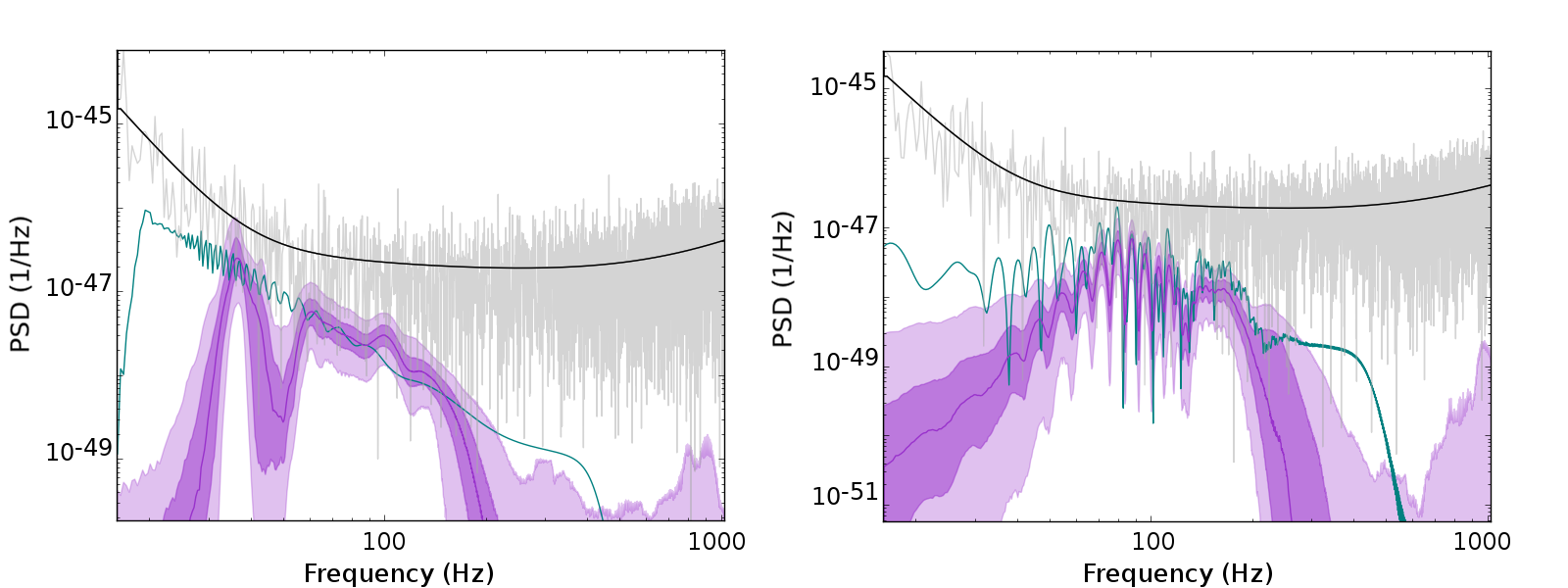}
\caption{The two upper plots show two injected \texttt{IMR} waveforms (plotted with green lines), i.e.\ the whitened (divided by the noise amplitude spectrum in the Fourier domain) strains as functions of time elapsed from the start of the binary evolution at 8 Hz orbital frequency, along with the medians of the signals reconstructed by BW (purple lines), the corresponding 50 and 90 percent intervals (dark and light purple areas, respectively), and the stationary, Gaussian noises with the embedded signals (gray lines). The waveform plotted in the left panels has an initial eccentricity of $e=0.1$, while the one plotted in the right panels has $e=0.7$. In the $e=0.7$ case, we can clearly see the repeated bursts before the merger.  The two lower plots show the power spectral densities (PSDs) of the different elements shown in the upper plots, each indicated with their respective colors; black lines show the median reconstructed PSDs of the noise with the embedded signal.  \label{fig:waveforms}}
\end{center}
\end{figure*}

Using each of the two waveform generators, and for each of the different $e$ values in the following set: $e=[10^{-6},\, 0.05,\, 0.10,\, 0.15,\, 0.20,\, 0.25,\, 0.30,\, 0.40,\, 0.50,\, 0.60,\, 0.70,\\ 0.80]$, we generated 500 BBH signals with $e$ initial eccentricity at 8 Hz initial orbital frequency, and embedded them in our simulated noise samples (this embedding process is conventionally referred to as \textit{signal injection} in GW research literature). Following previous works (see e.g.\ \citealt{2019ApJ...883..149A}), for this analysis we simulated BHs with zero spins. Figure \ref{fig:waveforms} shows two of such waveforms (one with $e=0.1$ and the other with $e=0.7$) along with their power spectral densities (PSDs), generated with the \texttt{IMR} algorithm. As the \texttt{EccentricTD} waveform generator cannot create signals with $e\geq 0.9$ reliably \citep{2016PhRvD..93f4031T} and as we are aware\footnote{From private communication with Sean T. McWilliams (Department of Physics and Astronomy, West Virginia University, Morgantown, West Virginia 26506, USA).} of other works in progress dealing with the $e\geq 0.9$ high-eccentricity regime, we only analyse signals with $e<0.9$ in this paper. We chose the masses of the BBH components independently from a uniform distribution within the range of $15 \ M_{\odot} \leq m_{1,2} \leq 25 \ M_{\odot}$, following \cite{2017ApJ...839...15B} and \cite{2015ApJ...800...81E}, who argued that the GW signals for these masses are compact in time-frequency space, making them ideal targets for generic burst searches. Also following the aforementioned two papers, we used a redshift distribution of the BBHs uniform in comoving volume between $10^{-4} \leq z \leq 0.33$. We chose all other extrinsic parameters (i.e.\ orbital inclination, coalescence phase, equatorial sky coordinates and polarization angle) from uniform distributions covering the full ranges of physically possible values.

To characterize the quality of waveform reconstruction, following \cite{2017ApJ...839...15B}, we use the $\mathcal{O}$ overlap, which measures the similarity of an injected $h_{\mathrm{i}}$ and a recovered $h$ signal waveform as:
\begin{equation}
    \mathcal{O} = \frac{(h_{\mathrm{i}} | h)}{\sqrt{(h_{\mathrm{i}} | h_{\mathrm{i}}) \cdot (h | h)}} ,
    \label{eq:O}
\end{equation}
where $(.|.)$ is a noise weighted inner product defined as:
\begin{equation}
    (a|b)=2\int_0^{\infty} \frac{a(f)b^*(f)+a^*(f)b(f)}{S_{\mathrm{n}}(f)}\,\mathrm{d}f,
\end{equation}
$S_{\mathrm{n}}$ is the one-sided power spectral density of the detector noise and $x^*$ denotes the complex conjugate of $x$. Eq. \ref{eq:O} defines $\mathcal{O}$ in a way that we get $\mathcal{O}=1$ in the case of a perfect match between $h_{\mathrm{i}}$ and $h$, $\mathcal{O}=0$ means no match, and $\mathcal{O}=-1$ means perfect anti-correlation. For a network of $N$ GW detectors, the waveform reconstruction can be characterized by the network overlap, calculated as follows:
\begin{equation}
    \mathcal{O}_{\mathrm{net}} = \frac{\sum_{j=1}^{N} (h_{\mathrm{i}}^{(j)} | h^{(j)})}{\sqrt{\sum_{j=1}^{N} (h_{\mathrm{i}}^{(j)} | h_{\mathrm{i}}^{(j)}) \cdot \sum_{j=1}^{N} (h^{(j)} | h^{(j)}) }},
\end{equation}
where $j$ denotes the $j$-th detector of the network. Since we used only H1 and L1 in our study, $N=2$ in our case. 

We also characterized the performance of BW in giving point estimates on the model-independent waveform central moments. We only carried out these tests for \texttt{IMR} waveforms. The first central moments are the central time ($t_0$) and the central frequency ($f_0$) of the signal, defined as:
\begin{equation}
    t_0 = \int_{-\infty}^{\infty} \varrho_{\mathrm{TD}}(t)\,t\,\mathrm{d}t,
\end{equation}
\begin{equation}
    f_0 = \int_{0}^{\infty} \varrho_{\mathrm{FD}}(f)\,f\,\mathrm{d}f
\end{equation}
where $\varrho_{\mathrm{TD}}$ and $\varrho_{\mathrm{FD}}$ denote the effective normalized distributions of signal energy expressed in the time-domain (TD) and in the frequency-domain (FD), respectively:
\begin{equation}
    \varrho_{\mathrm{TD}}(t)=\frac{h(t)^2}{h_{\mathrm{rss}}^2},
\end{equation}
\begin{equation}
    \varrho_{\mathrm{FD}}(f)=\frac{2|\tilde{h}(f)|^2}{h_{\mathrm{rss}}^2},
\end{equation}
$h(t)$ is the whitened (i.e.\ normalized by the noise amplitude spectrum in the Fourier domain) waveform for a given detector, $\tilde{h}(f)$ is its Fourier transform, and $h_{\mathrm{rss}}$ is the root sum squared strain amplitude parameter defined as $h_{\mathrm{rss}}^2=\int h(t)^2\, \mathrm{d}t$. The second central moments are the duration ($\Delta t$) and the bandwidth ($\Delta f$) of the signal, defined as:
\begin{equation}
    (\Delta t)^2 = \int_{-\infty}^{\infty} \varrho_{\mathrm{TD}}(t)\,(t-t_0)^2\,\mathrm{d}t,
\end{equation}
\begin{equation}
    (\Delta f)^2 = \int_{0}^{\infty} \varrho_{\mathrm{FD}}(f)\,(f-f_0)^2\,\mathrm{d}f.
\end{equation}
Following the analysis in \cite{2017ApJ...839...15B}, we calculated the median of the waveform moments for the samples in the Markov chain in order to give a point estimate of them, and we quantify the accuracy of these point estimates using the absolute error ($e_x$) as well as the relative error ($\eta_x$) of the estimation, defined as:
\begin{equation}
    e_x = |x^{(\mathrm{e})} - x^{(\mathrm{r})}|,
\end{equation}
\begin{equation}
    \eta_x = \frac{e_x}{x^{(\mathrm{r})}},
\end{equation}
where $x^{(\mathrm{e})}$ is the estimated and $x^{(\mathrm{r})}$ is the real value of $x$. All moments are calculated for the H1 detector data, however, by using the L1 detector data we get very similar results.

\section{Results} \label{sec:results}

In this section we show how well BW reconstructs $e>0$ BBH signals in terms of network overlap, the percentage of injected waveforms identified by BW as signals, and the errors of the central moment estimations given by BW (see Section \ref{sec:reconstruction}). We also compare the results we obtained using the sine-Gaussian wavelet base in BW to those we obtained with the chirplet base (see Section \ref{sec:chirplet}). Note that the results we present here only correspond to signals for which the Bayes factor calculated by BW was in favor of the signal model (see \citealt{2015CQGra..32m5012C} for more details).

\subsection{Waveform reconstruction} \label{sec:reconstruction}

\begin{figure}
\begin{center}
\includegraphics[scale=0.5,angle=0]{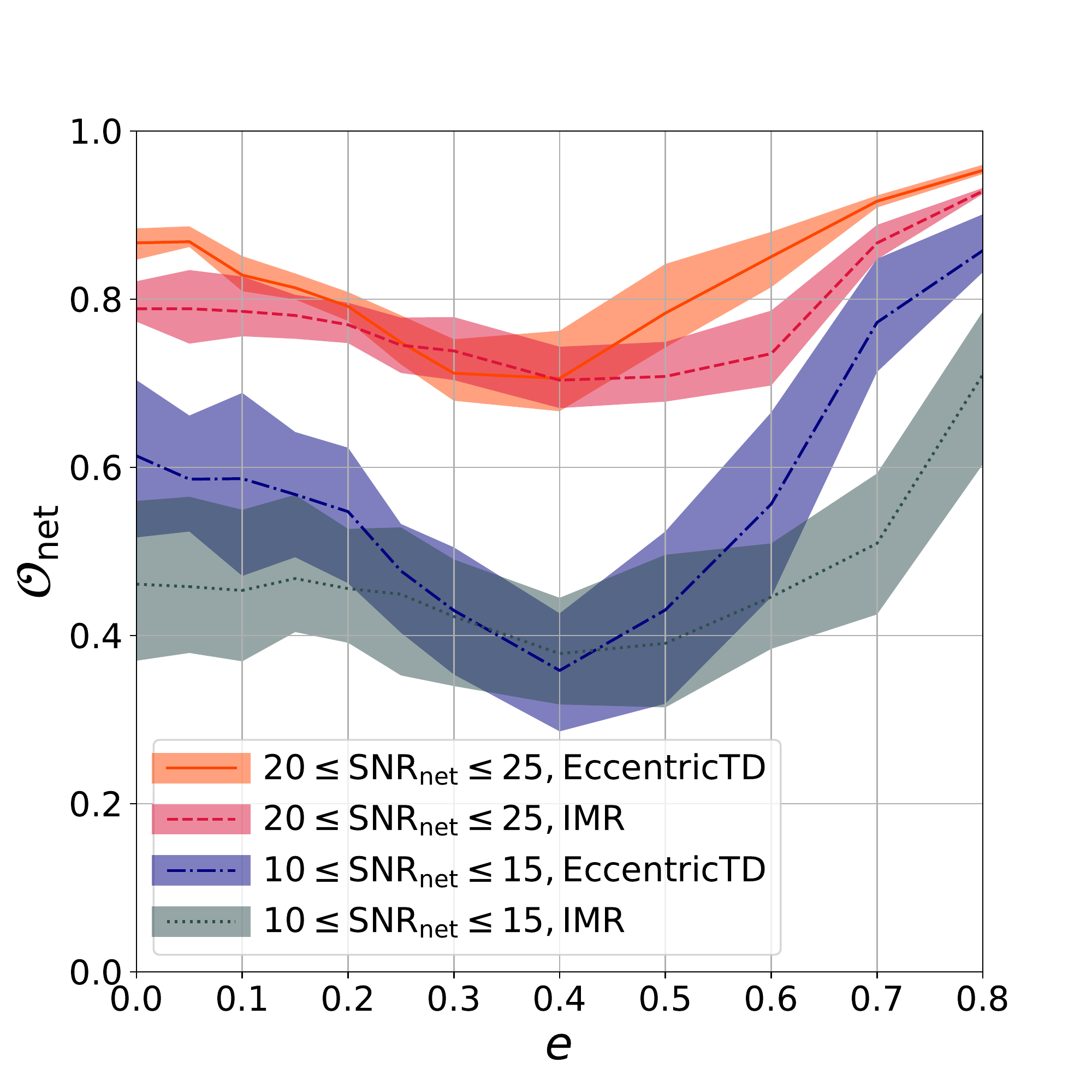}
\caption{Dependence of the median values of network overlaps ($\mathcal{O}_{\mathrm{net}}$) on initial eccenticities ($e$) for signals generated with the \texttt{EccentricTD} and the \texttt{IMR} algorithms. For both types of simulated waveforms we show results in two network signal-to-noise ratio (SNR$_{\mathrm{net}}$) ranges. Shaded areas represent the regions between the $20^{\mathrm{th}}$ and $80^{\mathrm{th}}$ percentiles of the measured $\mathcal{O}_{\mathrm{net}}$ values. For all four cases we can see a clear trend of $\mathcal{O}_{\mathrm{net}}$ decreasing with $e$ up until $e\simeq 0.4$, where the trend changes to $\mathcal{O}_{\mathrm{net}}$ monotonously increasing with $e$. \label{fig:eccplot}}
\end{center}
\end{figure}

\begin{figure}
\begin{center}
\includegraphics[scale=0.5,angle=0]{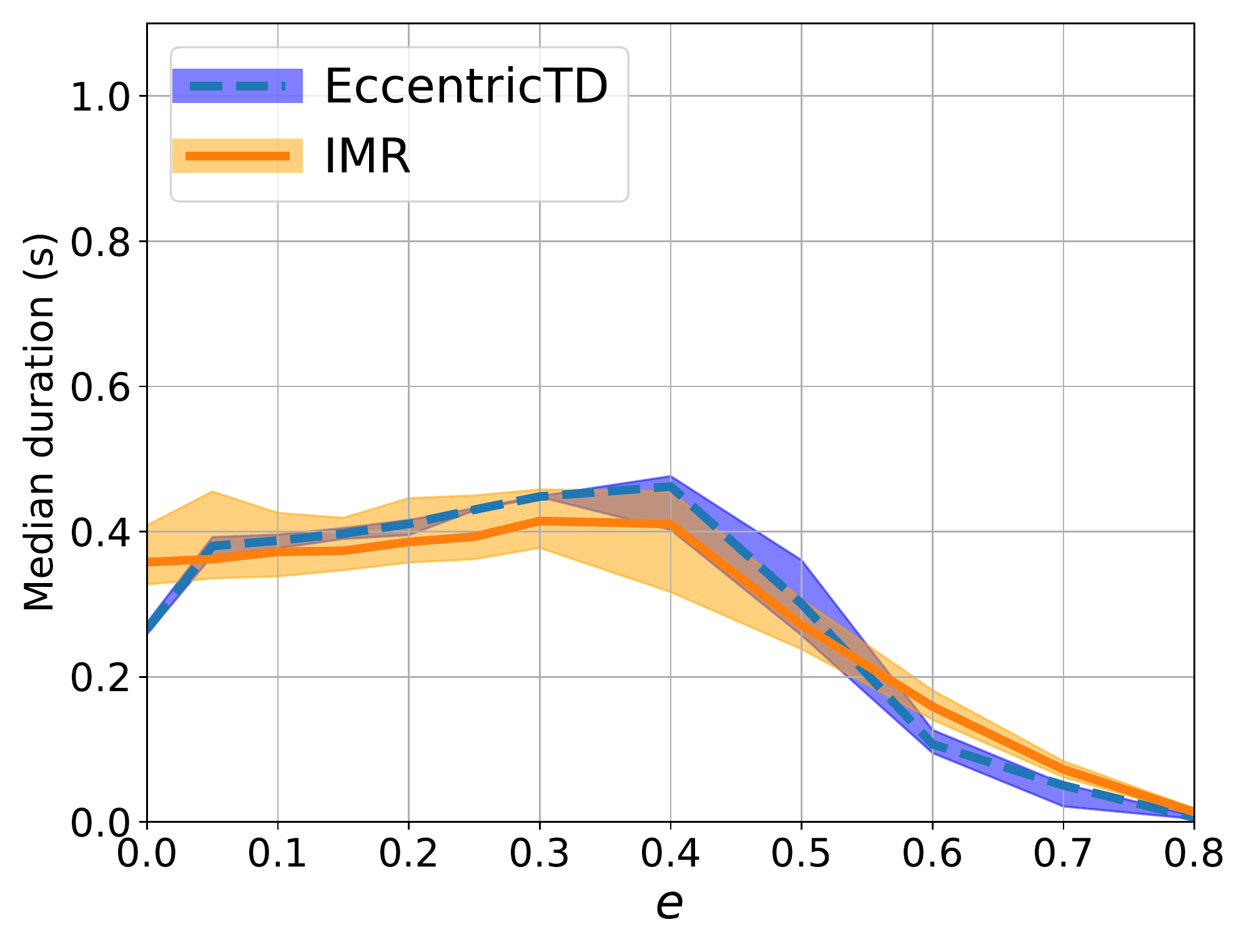}
\caption{The median durations of the injected signals with initial eccentricity $e$. Shaded areas represent the regions between the $20^{\mathrm{th}}$ and $80^{\mathrm{th}}$ percentiles. The duration of the signals decreases with $e$ above $e \geq 0.4$.  \label{fig:durations}}
\end{center}
\end{figure}

\begin{figure}
\begin{center}
\includegraphics[scale=0.5,angle=0]{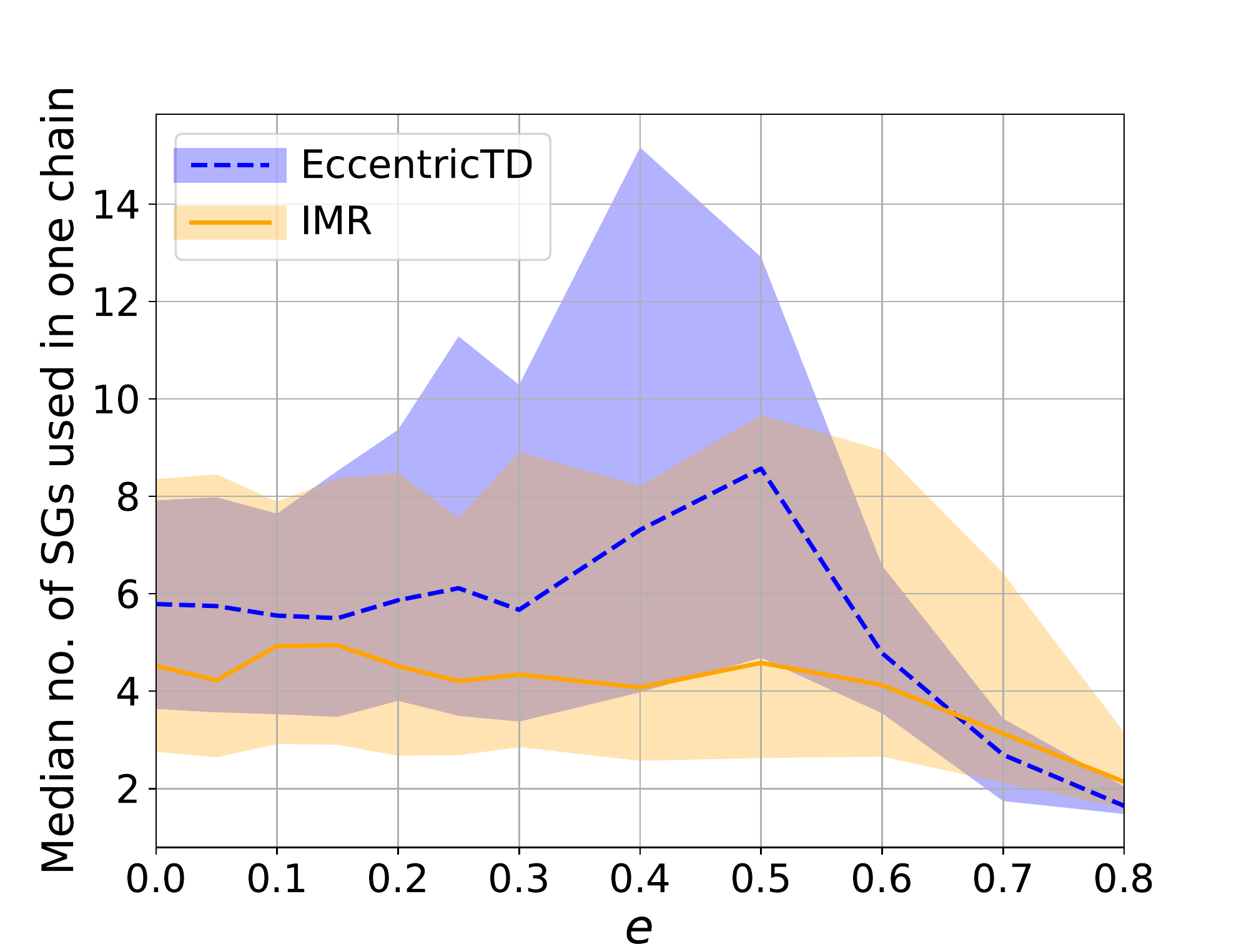}
\caption{The median number of sine-Gaussian wavelets used by BW in one Markov chain to reconstruct the injected signal with initial eccentricity $e$. Shaded areas represent the regions between the $20^{\mathrm{th}}$ and $80^{\mathrm{th}}$ percentiles. We show results for both waveforms generated by the \texttt{EccentricTD} and the \texttt{IMR} algorithms. The number of wavelets used decreases significantly for $e\gtrsim0.5$. The reason behind this is that while keeping all other parameters equal, signals with higher initial eccentricities are shorter, furthermore, these signals consist of repeated distinct bursts, which can be reconstructed with distinct wavelets. \label{fig:noofSGs}}
\end{center}
\end{figure}

Figure \ref{fig:eccplot} shows the dependence of $\mathcal{O}_{\mathrm{net}}$ network overlaps on the $e$ initial eccentricities of BBHs. We show results for signals generated with both the \texttt{EccentricTD} and the \texttt{IMR} waveform generators, and for two different network signal-to-noise ratio ($\mathrm{SNR}_{\mathrm{net}}$) ranges. We define $\mathrm{SNR}_{\mathrm{net}}$ as the root sum squared signal-to-noise ratios of the two aLIGO detectors:
\begin{equation}
    \mathrm{SNR}_{\mathrm{net}} = \sqrt{\mathrm{SNR}_{\mathrm{L1}}^2+\mathrm{SNR}_{\mathrm{H1}}^2},
\end{equation}
where the signal-to-noise ratio is defined as:
\begin{equation}
    \mathrm{SNR} = 2\,\sqrt{\int_0^{\infty} \frac{|\tilde{h}(f)|^2}{S_{\mathrm{n}} (f)}\,\mathrm{d}f}.
\end{equation}
The shaded areas in Figure \ref{fig:eccplot} represent the regions between the $20^{\mathrm{th}}$ and $80^{\mathrm{th}}$ percentiles of the measured $\mathcal{O}_{\mathrm{net}}$ distributions. As expected, signals in the higher $\mathrm{SNR}_{\mathrm{net}}$ band are reconstructed with higher $\mathcal{O}_{\mathrm{net}}$ values. For all four cases we can see a clear trend of decreasing $\mathcal{O}_{\mathrm{net}}$ as a function of $e$ up until $e\simeq 0.4$, and then a monotonous increase in $\mathcal{O}_{\mathrm{net}}$ with $e$. The reason behind the increasing network overlap for $e \gtrsim 0.4$ lies in the fact that signals become shorter in time as $e$ increases (see Figure \ref{fig:durations}), and therefore BW can reconstruct signals with less number of sine-Gaussian wavelets (see the curves in Figure \ref{fig:noofSGs} above $e\simeq0.5$ for a confirmation), usually resulting in higher $\mathcal{O}_{\mathrm{net}}$ values for the reconstructions. 

\begin{figure}
\begin{center}
\includegraphics[scale=0.5,angle=0]{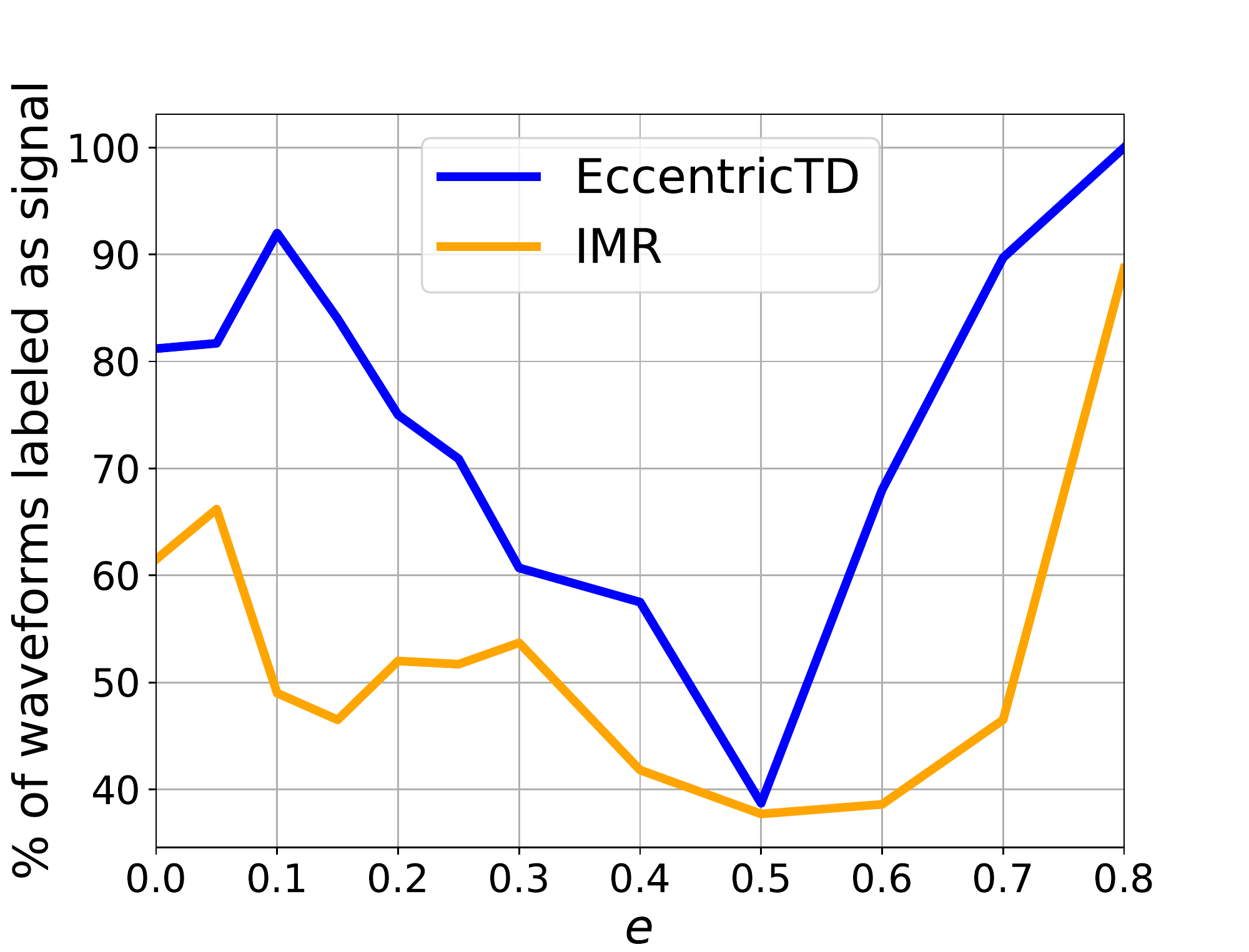}
\caption{The percentage of waveforms within the range $10 \leq \mathrm{SNR_{net}} \leq 15$ with initial eccentricity $e$ for which the Bayes factor calculated by BW is in favor of the signal model. For the discussion of the results plotted here, see Section \ref{sec:reconstruction}.} \label{fig:percofrec}
\end{center}
\end{figure}

Figure \ref{fig:percofrec} shows the percentage of injected waveforms within the range $10 \leq \mathrm{SNR_{net}} \leq 15$ with initial eccentricity $e$ for which the Bayes factor calculated by BW is in favor of the signal model. We have chosen to show the results only in a narrow $\mathrm{SNR_{net}}$ range in order to mitigate the effect of varying median $\mathrm{SNR_{net}}$ values for different values of $e$. The figure shows that for $e \geq 0.5$ the percentage of signals correctly labeled as signals increases. The reason behind this is that as Figure \ref{fig:eccplot} shows these signals are reconstructed with a higher $\mathcal{O}_{\mathrm{net}}$ what makes their classification easier.

\begin{figure}
\begin{center}
\includegraphics[scale=0.5,angle=0]{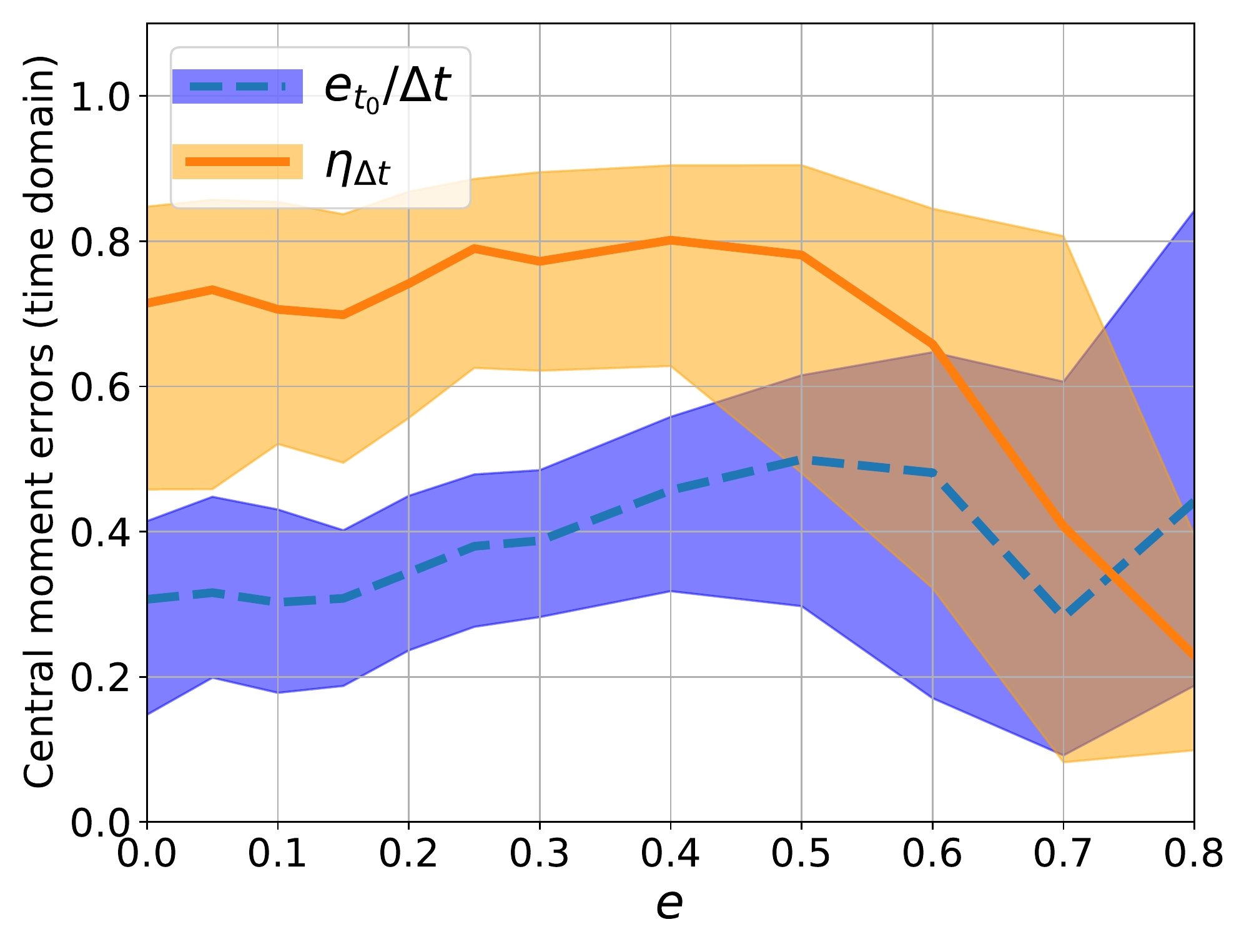}
\includegraphics[scale=0.5,angle=0]{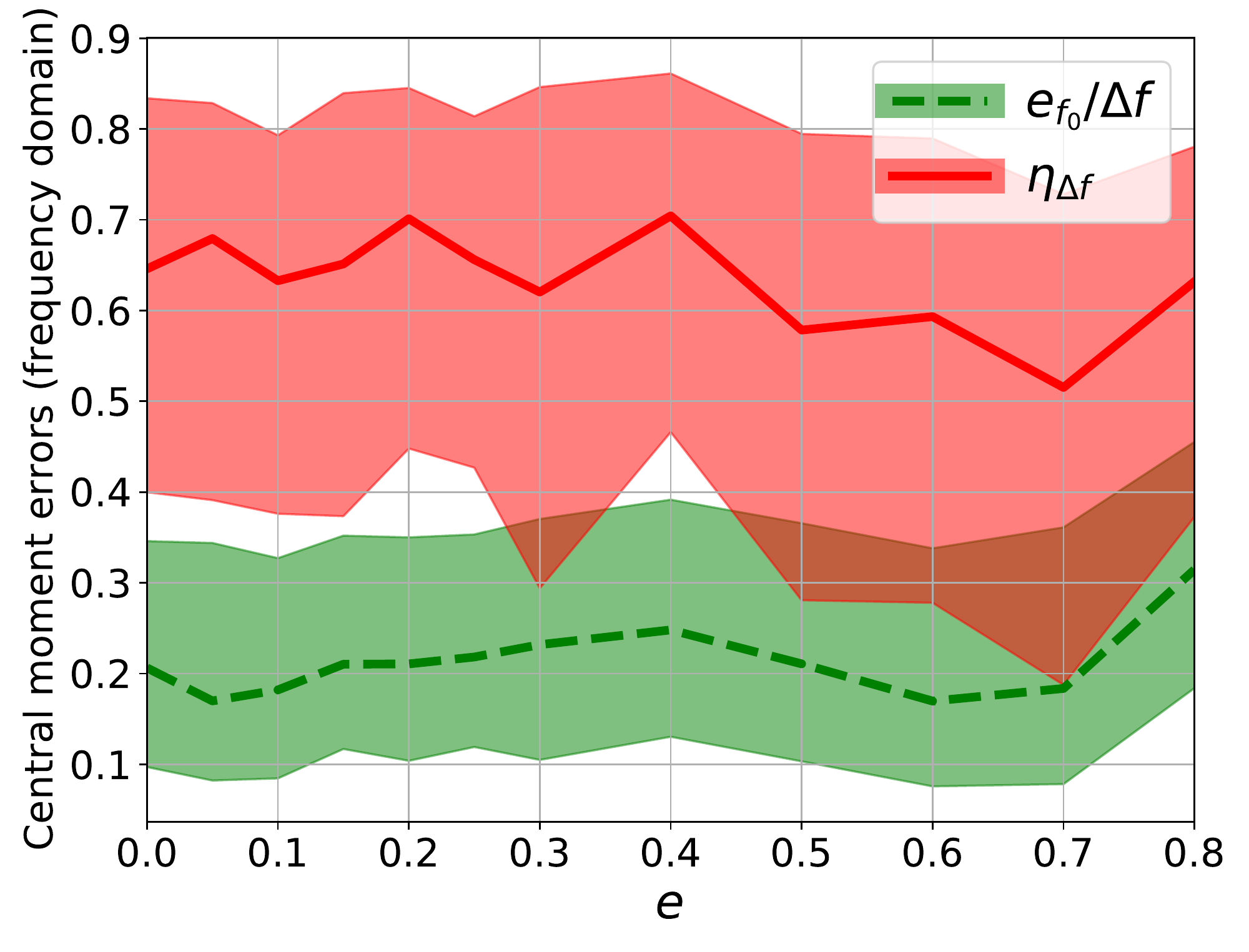}
\caption{Median values of \texttt{IMR} waveform central moment errors for signals with different initial eccentricities. The upper panel shows the central moment errors in the time domain, while the lower panel shows the errors in the frequency domain. Shaded areas represent the values between the 25$^{\mathrm{th}}$ and the 75$^{\mathrm{th}}$ percentiles. For the discussion of the results plotted here, see Section \ref{sec:reconstruction}. \label{fig:modelind}}
\end{center}
\end{figure}

 Figure \ref{fig:modelind} shows the dependence of the waveform central moment errors ($e_{t_0}/\Delta t$, $e_{f_0}/\Delta f$, $\eta_{\Delta t}$ and $\eta_{\Delta f}$) on the initial eccentricity $e$, showing the results of tests we only carried out for \texttt{IMR} waveforms due to limited amount of computational resources. We divided the absolute errors of the first moment estimations ($e_{t_0}$ and $e_{f_0}$) with the real values of the corresponding second moments ($\Delta t$ and $\Delta f$) because we expect that statistical errors of first moment estimations scale with the real values of these second moments. In the upper panel of Figure \ref{fig:modelind}, we can see that the median of $\eta_{\Delta t}$ values is approximately constant for $e\lesssim0.5$, and starts decreasing for higher eccentricities. The reason behind this is that the waveform for these eccentricities consists of distinct and repeated bursts (see the right panel in Figure \ref{fig:waveforms} for an example). In order to give an accurate estimate on the signal duration, BW needs to reconstruct the full waveform of the signal, including the low-amplitude beginning of it as well. When the signal consists of distinct bursts, even the first of such bursts can have a high enough amplitude for BW to reconstruct it effectively. However it is more difficult for BW to reconstruct the beginning of a small-eccentricity signal, where the first burst has lower amplitude and is less articulated. This makes the estimation of the duration less accurate for signals with $e\lesssim0.5$, as can be seen on the orange solid curve in the upper plot of Figure \ref{fig:modelind}. As $e$ increases, the estimation of the central time becomes more accurate as well. However, since signals with higher initial eccentricities are shorter (when all other parameters are kept the same), there is an increase in $e_{t_0}/\Delta t$ as $e$ increases (see the blue dashed line in the upper plot of Figure \ref{fig:modelind}).

The lower panel of Figure \ref{fig:modelind} shows that the central moment errors in the frequency domain remain roughly the same with increasing initial eccentricity. For higher $e$ values the bandwidths of signals increase due to the presence of higher-order harmonics in GW radiation, but as $e_{f_0}$ also increases for higher initial eccentricities, the ratio of the two, $e_{f_0}/\Delta f$, remains nearly constant. Note that the results are similar for $\eta_{\Delta f}$ as well. 

\subsection{Comparison of the sine-Gaussian and chirplet bases} \label{sec:chirplet}

As generic GW signals have frequency content that evolves in time, it is expected that a collection of chirplets (i.e.\ generalized wavelets with linear frequency evolution) provides a more compact representation of signals than a collection of sine-Gaussian wavelets \citep{2018PhRvD..97j4057M}. As \cite{2018PhRvD..97j4057M} has shown, this more compact representation results in more accurate waveform reconstruction, especially for events with low SNR$_{\mathrm{net}}$ and those that occupy a large volume in time-frequency space. The authors of \cite{2018PhRvD..97j4057M} have investigated how using a chirplet base improves the performance of BW in the reconstruction of BBH signals with zero eccentricity, as well as of unpolarized white noise bursts. In this section, we have extended the investigation to BBH signals with nonzero eccentricities.

\begin{figure}
\begin{center}
\includegraphics[scale=0.5,angle=0]{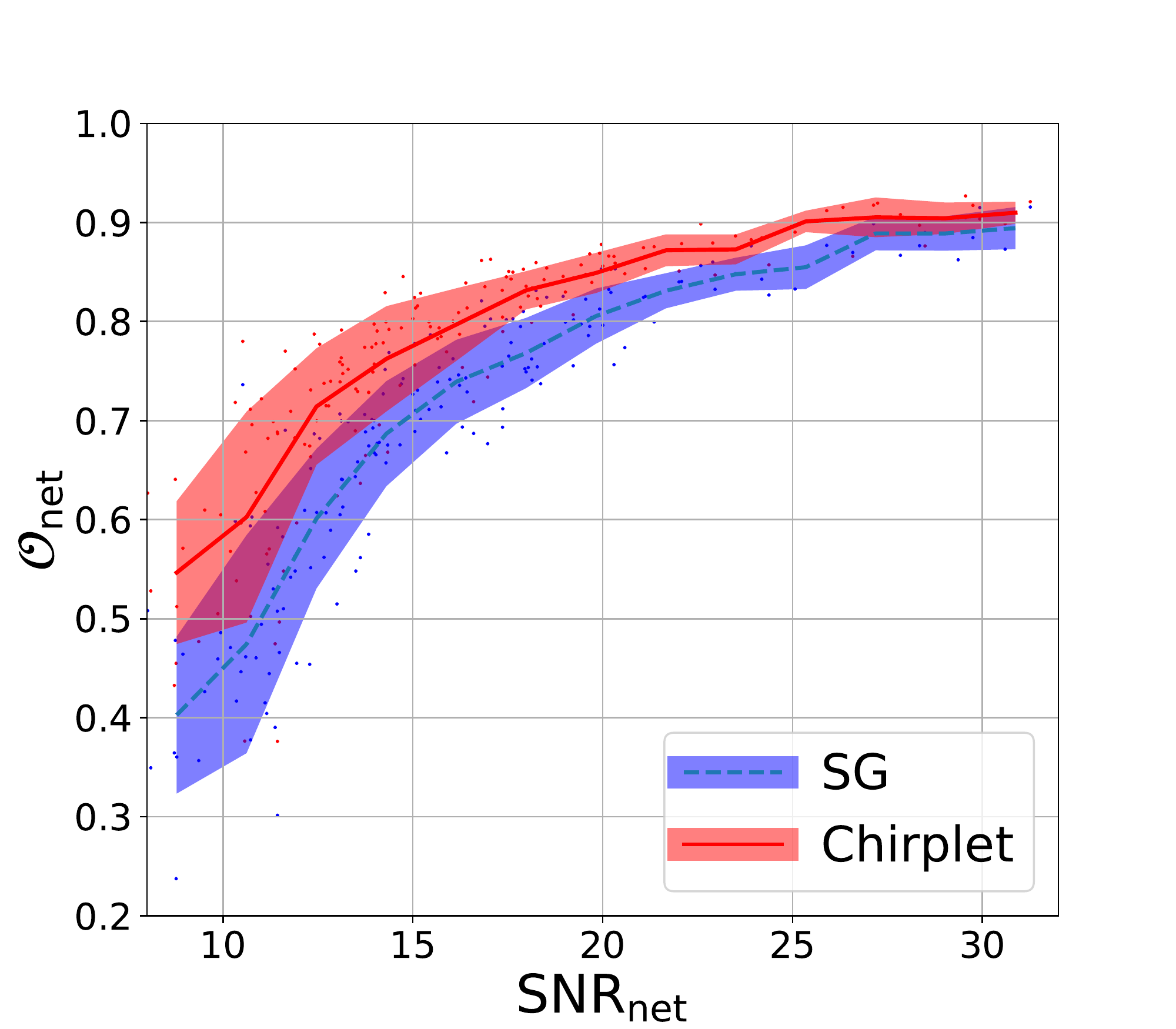}
\caption{The network overlap as a function of the network signal-to-noise ratio for \texttt{IMR} waveforms with $e=0.1$ initial eccentricity. Blue and red dots represent waveforms reconstructed using the sine-Gaussian and the chirplet base, respectively. The blue and red lines show the mean of the network overlaps, while the shaded areas represent the 1$\sigma$ uncertainty regions. \label{fig:chhist}}
\end{center}
\end{figure}

\begin{figure}
\begin{center}
\includegraphics[scale=0.5,angle=0]{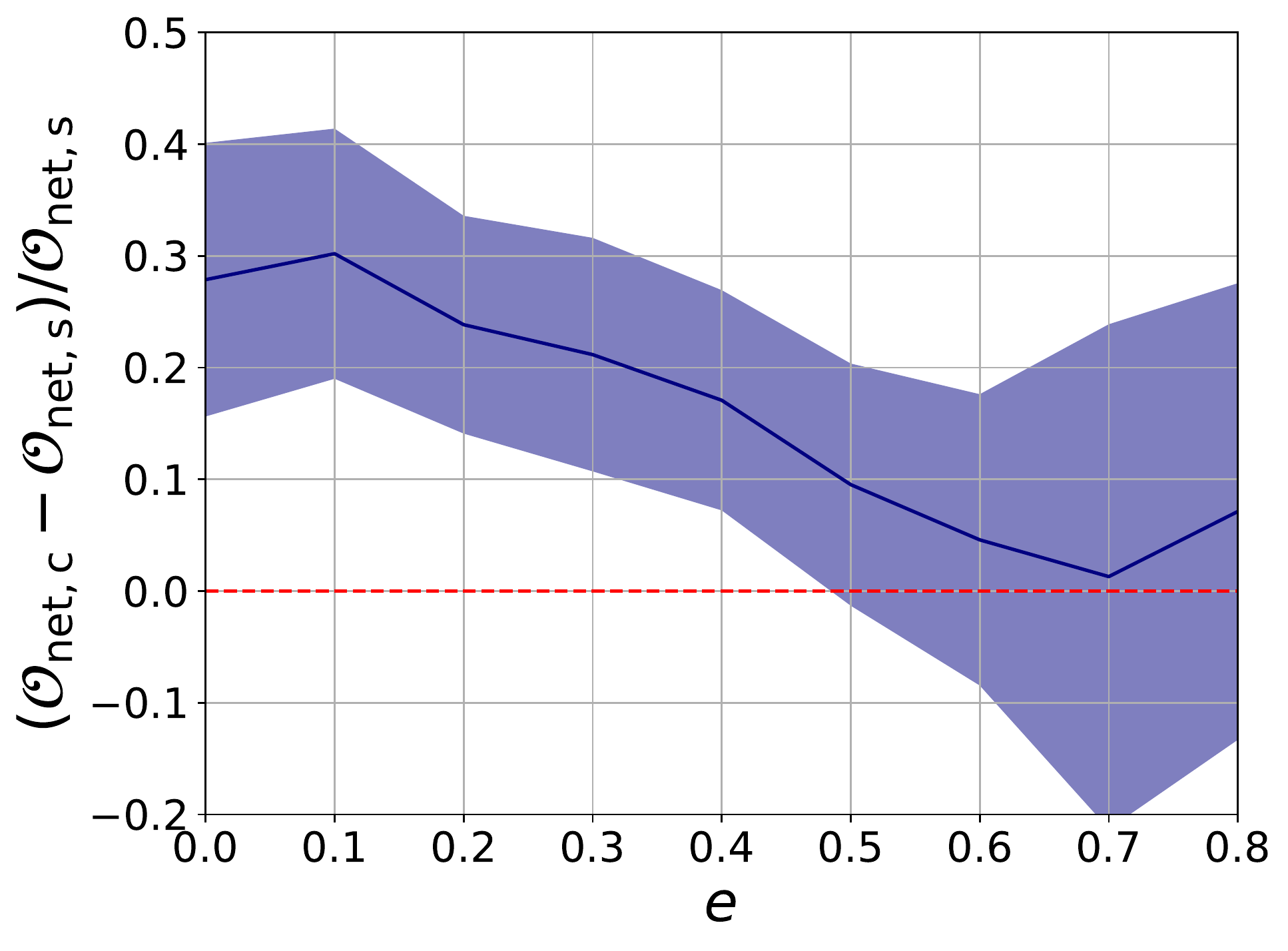}
\caption{The relative improvement in network overlap of \texttt{IMR} waveform reconstructions as a function of initial eccentricity $e$ when using chirplet base, compared to when using sine-Gaussian wavelet base. In ($\mathcal{O}_{\mathrm{net, c}} - \mathcal{O}_{\mathrm{net, s}}) / \mathcal{O}_{\mathrm{net, s}}$, $\mathcal{O}_{\mathrm{net, c}}$ is the network overlap when using sine-Gaussian base, and $\mathcal{O}_{\mathrm{net, s}}$ is the network overlap when using chirplet base. The shaded area represents the 1$\sigma$ uncertainty region, and the dashed line shows the case of zero difference between the two cases. \label{fig:chdiff}}
\end{center}
\end{figure}

In order to test how the usage of chirplets improves the reconstruction of BBH signals, we injected \texttt{IMR} waveforms to noise samples as described in Section \ref{sec:reconstruction}, and reconstructed these signals using both the sine-Gaussian and the chirplet base. Figure \ref{fig:chhist} shows the mean network overlap as a function of the network signal-to-noise ratio for both cases, using signals with $e=0.1$ initial eccentricities. Note, that the general behaviour of these curves for other $e$ values is very similar to the one shown in Figure \ref{fig:chhist}. Similarly to the results of \cite{2018PhRvD..97j4057M} for $e=0$, there is significant increase in $\mathcal{O}_{\mathrm{net}}$ as SNR$_{\mathrm{net}}$ increases up to SNR$_{\mathrm{net}} \lesssim 25$, and for higher SNR$_{\mathrm{net}}$ values the obtained overlaps remain at a comparable ($\mathcal{O}_{\mathrm{net}}\simeq 0.9)$ level.

Figure \ref{fig:chdiff} shows the relative improvement in network overlap when using chirplet base compared to the case when using sine-Gaussian base. The figure shows that the increase in performance is more significant for lower initial eccentricities, which is consistent with the finding of \cite{2018PhRvD..97j4057M}, namely that the improvement is more significant for signals that occupy a large volume in time-frequency space.

\begin{figure}
\begin{center}
\includegraphics[scale=0.5,angle=0]{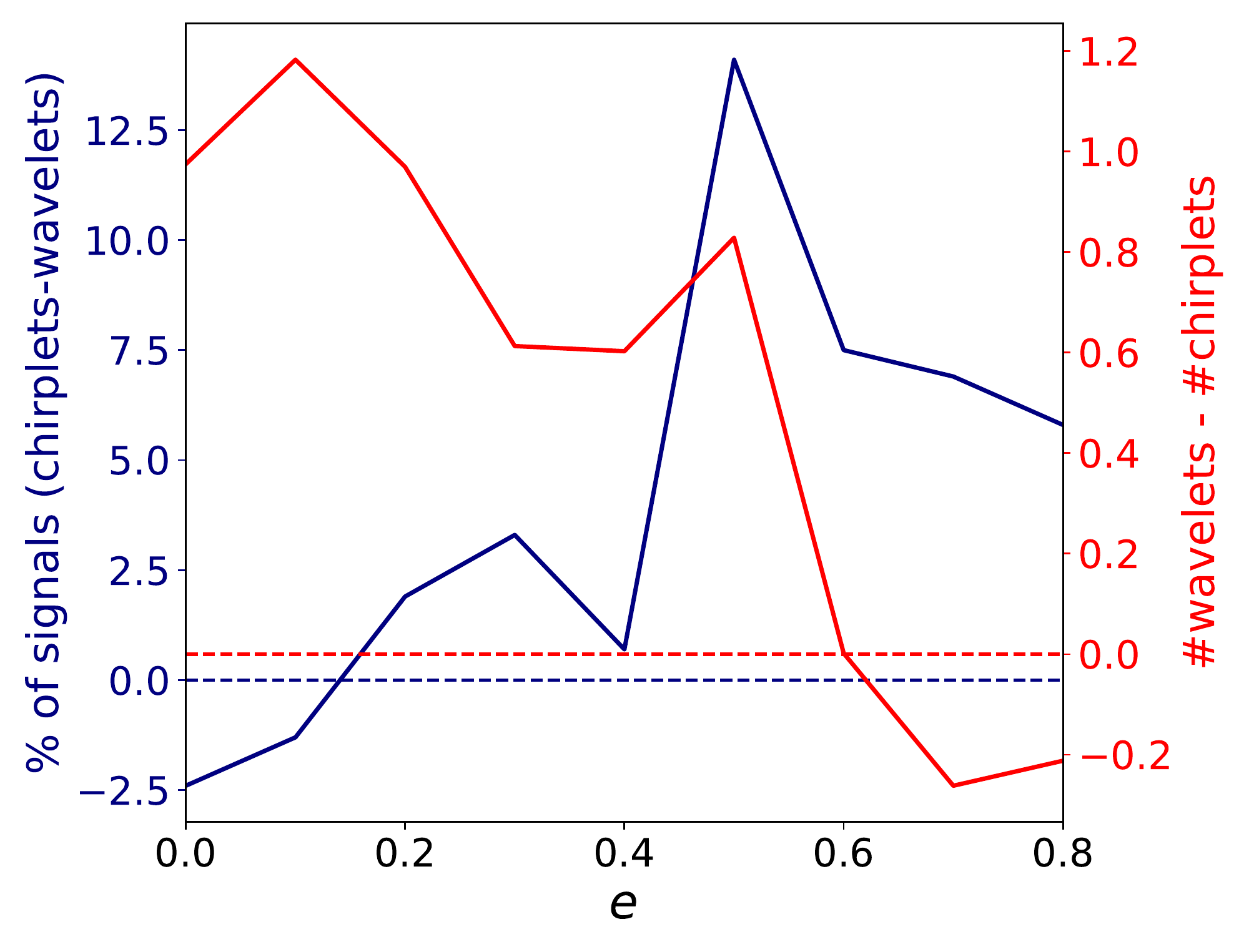}
\caption{The blue line shows the improvement in correctly identifying signals when using the chirplet base, i.e.\ the percentage of injected waveforms with initial eccentricity $e$ for which the Bayes factor was in favor of the signal model when using chirplets minus the percentage when using the sine-Gaussian wavelet base. The red line shows the difference between the median number of sine-Gaussian wavelets and chirplets used by BW in one Markov chain to reconstruct the injected signal with initial eccentricity $e$. \label{fig:noSGch}}
\end{center}
\end{figure}

The blue line in Figure \ref{fig:noSGch} shows the improvement in BW correctly classifying reconstructions as signals when using the chirplet base, i.e.\ the difference in the percentage of injected \texttt{IMR} waveforms with initial eccentricity $e$ for which the Bayes factor was in favor of the signal model when using the chirplet base and when using the sine-Gaussian wavelet base. Note that when the chirplet base is used, for signals with $e\gtrsim 0.5$ there is a $\sim10\%$ improvement in the number of reconstructions correctly classified by BW as signals. The red line in Figure \ref{fig:noSGch} shows the difference between the median number of sine-Gaussian wavelets and chirplets used by BW in one Markov chain to reconstruct the injected signal with initial eccentricity $e$. For signals with $e\lesssim 0.5$ chirplets provide more compact representations of signals, i.e.\ less chirplets than sine-Gaussian wavelets are needed to reconstruct injected signals. This difference vanishes for signals with higher ($e\gtrsim0.6$) initial eccentricities.

We also investigated how the median values of the \texttt{IMR} waveform central moment errors change when we use the chirplet base instead of the sine-Gaussian base. Although there seemed to be a slight improvement in the median of $\eta_{\Delta a}$ for signals with $e\lesssim 0.25$, the errors are too large to say anything conclusive about it. There was no significant change in the values of the other central moment errors.

\section{Conclusion} \label{sec:conclusion}

We have presented a comprehensive study on the performance of BayesWave (a Bayesian waveform reconstruction and parameter estimation tool used by the LIGO-Virgo Collaboration) in reconstructing GW signals of BBH systems with nonzero eccentricities. We generated simulated signals with initial eccentricities between $e=10^{-6}$ and $e=0.8$ at 8 Hz orbital frequencies using two different waveform simulators (\texttt{EccentricTD} and \texttt{IMR}) and injected them to simulated stationary, Gaussian noise samples mocking the predicted noise of design sensitivity aLIGO.

We characterized the goodness of waveform reconstruction by the network overlap ($\mathcal{O}_{\mathrm{net}}$). For higher network signal-to-noise ratios (SNR$_{\mathrm{net}}$) we get better reconstructions for both \texttt{EccentricTD} and \texttt{IMR} waveforms, independently from $e$. For signals in a given SNR$_{\mathrm{net}}$ range, there is a clear trend of decreasing $\mathcal{O}_{\mathrm{net}}$ as a function of the initial eccentricity up until $e \simeq 0.4$ and afterwards we experience a monotonous increase. The reason for this increase lies in the fact that the signals get shorter and BW can reconstruct them with less number of sine-Gaussian wavelets. Due to the higher $\mathcal{O}_{\mathrm{net}}$, within a given SNR$_{\mathrm{net}}$ range BW classifies more signals correctly in this $e$ range.

We have also tested how accurately BW can estimate the central moments of BBH waveforms with nonzero eccentricity. These central moments are model-independent parameters of a signal, therefore by examining the estimation of them we can characterize the parameter estimation capabilities of BW without assuming any astrophysical model for the source. The estimates for the absolute error of the central frequency divided by the bandwidth ($e_{f_0}/\Delta f$) and for the relative error of the bandwidth ($\eta_{\Delta f}$) are nearly independent of the initial eccentricity, while the error of the central time divided by the duration ($e_{t_0}/\Delta t$) increases for higher $e$ values and the relative error of the duration ($\eta_{\Delta t}$) decreases for signals with $e\gtrsim 0.5$.

BayesWave can reconstruct generic transient signals with sine-Gaussian wavelets or using generalized wavelets with linear frequency evolution (chirplets). In this paper we have also quantified various aspects of the difference of the two reconstruction methods when applied to BBH signals with nonzero eccentricities. BW reconstructs signals with higher $\mathcal{O}_{\mathrm{net}}$ using chirplets, especially for lower SNR$_{\mathrm{net}}$ values and lower eccentricities. For signals with $e\gtrsim 0.5$ there is a $\sim10\%$ improvement in correctly classifying reconstructions as signals when the chirplet base is used. For signals with $e\lesssim 0.5$ less number of chirplets are needed for the reconstruction compared to the number of wavelets needed when the sine-Gaussian base is used.

Based on our findings, BW can be an effective tool to reconstruct GWs from BBHs with nonzero eccentricities and estimate their model-independent signal parameters, especially for signals with $e\lesssim0.2$ or $e\gtrsim0.7$. By using chirplets instead of sine-Gaussian wavelets, BW's accuracy in waveform reconstruction can significantly be improved for signals with $e\lesssim0.6$.\\

\section*{Acknowledgements}

This paper was reviewed by the LIGO Scientific Collaboration under LIGO Document P2000173. We thank Imre Bartos, Neil Cornish, James Clark, Sudarshan Ghonge, Bence Kocsis, Margaret Millhouse, J\'anos Tak\'atsy and Shubhanshu Tiwari for their valuable help in the project. The authors thank Jonah Kanner and Paul Lasky for their useful comments on the manuscript. The authors are grateful for computational resources provided by the LIGO Laboratory and supported by National Science Foundation Grants PHY-0757058 and PHY-0823459. Gergely D\'alya was supported by the \'UNKP-18-3 New National Excellence Program of the Ministry of Human Capacities.

\end{document}